\documentclass[aps,pra,twocolumn,superscriptaddress,notitlepage,showpacs,showkeys]{revtex4-1}
\usepackage{graphicx,amsmath,amsfonts,amssymb,upgreek,txfonts,color}
\usepackage[colorlinks,linkcolor=blue,citecolor=blue,urlcolor=blue,breaklinks=true]{hyperref}
\usepackage{color, colortbl}
\definecolor{lightgreen}{rgb}{0.88,1,1}

\usepackage[english]{babel}

\begin{document}



\title{Parity-time symmetry-enhanced simultaneous magnon and photon blockade in cavity magnonic system }

 \author{M. S. \surname{Ebrahimi} }
 \email{mehrosadatebrahimi2016@gmail.com}
 \address{Department of Physics, Faculty of Science, University of Isfahan, Hezar Jerib, 81746-73441, Isfahan, Iran}

\author{M. Bagheri Harouni}
 \email{m.bagheri@sci.ui.ac.ir}
 \address{Department of Physics, Faculty of Science, University of Isfahan, Hezar Jerib, 81746-73441, Isfahan, Iran}
 \address{Quantum Optics Group, Department of Physics, Faculty of Science, University of Isfahan, Hezar Jerib, 81746-73441, Isfahan, Iran}


 \date{\today}

 \begin{abstract}
 The main challenge in the recent demonstration of conventional magnon blockade is to increase the nonlinearity of the system especially in comparison with the dissipation channels. One can consider the Kerr nonlinearity through which magnon blockade in a cavity magnonic system is possible provided that the Kerr nonlinearity  is much stronger than the cavity and magnon mode dissipation rates.  In the present contribution, we consider a  parity-time ($\mathcal{P}\mathcal{T}$)-symmetric cavity magnonic system and study the effect of $\mathcal{P}\mathcal{T}$-symmetry phase on the magnon statistics and hence magnon blockade. We show that the $\mathcal{P}\mathcal{T}$-symmetry phase, which is achievable  by properly selecting the system parameters, can relax the requirement of large Kerr nonlinearity such that a perfect magnon blockade can be easily obtained even under a small value of Kerr nonlinearity. Surprisingly, although there is no photonic Kerr nonlinearity in the scheme, photon blockade can also occur simultaneously with magnon blockade. This result is arising from the $\mathcal{P}\mathcal{T}$-symmetric phase which can generate an effective photonic Kerr nonlinearity.

 \end{abstract}

\maketitle


\section{Introduction}\label{sec1}
In the last decade, hybrid magnonic systems based on spin wave collective excitations in ferromagnetic materials, like yttrium iron garnet (YIG) with very high spin density and low magnon dissipation rate have attracted a great deal of attention \cite{hybrid magnonic system 1, hybrid magnonic system 2, hybrid magnonic system 3, hybrid magnonic system 4, hybrid magnonic system 5, hybrid magnonic system 6, hybrid magnonic system 7, hybrid magnonic system 8}. 
The prominent feature of magnons, the ability of coherent interaction with optical and microwave photons \cite{hybrid magnonic system 1, hybrid magnonic system 2, hybrid magnonic system 3, hybrid magnonic system 4, hybrid magnonic system 5, hybrid magnonic system 6, hybrid magnonic system 7, hybrid magnonic system 8}, phonons \cite{cavity magnomechanics}, and superconducting qubits \cite{magnon meets the superconducting qubit} is providing a vast novel platforms for applications in quantum technologies such as quantum information processing \cite{entanglement 1, entanglement 2, entanglement 3, entanglement 4, entanglement 5, entanglement 6}, and quantum sensing \cite{PT symmetry magnetometry, Single-quadrature magnetometry}. 
Magnons, the elementary excitations of magnetically ordered systems, which their interaction with microwave photons in the strong and even ultrastrong coupling regimes has been demonstrated \cite{hybrid magnonic system 1, hybrid magnonic system 2}, can have long lifetimes and long coherence times \cite{YIG magnonics, Magnon spintronics, Spin waves}, which makes them suitable for different applications in quantum information. 
In the last decade, different quantum behaviour of magnons such as magnon-magnon or magnon-photon-phonon entanglement \cite{entanglement 1, entanglement 2, entanglement 3, entanglement 4, entanglement 5, entanglement 6}, magnon and phonon squeezing \cite{Squeezed states of magnons}, bell state generation \cite{bell state 1, bell state 2, bell state 3} as well as magnon blockade \cite{magnon blockade 1, magnon blockade 2,  magnon blockade 3,  magnon blockade 4,  magnon blockade 5}, as a pure quantum phenomenon have been theoretically studied. Magnon blockade, which similar to photon and phonon blockade in cavity quantum electrodynamics \cite{photon blockade CQED 1} and cavity optomechanics \cite{photon blockade opto 1}, is essentially originated from the anhamonicity in energy eigenvalues of the system,  has been theoretically proposed in various hybrid magnonics systems \cite{magnon blockade 1, magnon blockade 2, magnon blockade 3, magnon blockade 4, magnon blockade 5}. By increasing the nonlinearity of the magnon mode and subsequently the anhamonicity of the energy level spacing, the second magnon excitation is blocked once the first one has been excited. This quantum characteristic can lead to a single-magnon source with nonclassical magnon antibunching statistics in which the excitation of two and multi magnons is suppressed. 
Evidently, in conventional magnon blockade as the anharmonicity of the eigenenergy spectrum is larger, the magnon blockade is stronger and a perfect magnon blockade requires a large value of magnon nonlinearity, such as magnon Kerr nonlinearity in the system.

On the other hand, recent years have witnessed of striking developing practical techniques from the $\mathcal{P}\mathcal{T}$ symmetry structures \cite{pt symmetry app 1, low power optical isolation 1, low power optical isolation 2, single mode microcavity laser 1, single mode microcavity laser 2, loss-induced or gain induced transparency 1, loss-induced or gain induced transparency 2}. $\mathcal{P}\mathcal{T}$ symmetric systems have non-Hermitian Hamiltonian and their prominent feature is that by properly selecting the system parameters, they can have purely real eigenvalues \cite{pt symmetry in optics}. Where the system lose the $\mathcal{P}\mathcal{T}$ symmetry, is known as exceptional point (EP) in which the pair of eigenvalues collide. Transition from an unbroken $\mathcal{P}\mathcal{T}$-symmetric phase to a spontaneous $\mathcal{P}\mathcal{T}$ symmetric broken phase has enabled applications such as low power optical isolation \cite{low power optical isolation 1, low power optical isolation 2}, single mode microcavity laser \cite{single mode microcavity laser 1, single mode microcavity laser 2}, loss-induced or gain induced transparency \cite{loss-induced or gain induced transparency 1, loss-induced or gain induced transparency 2}, enhanced photon blockade \cite{PT symmetry enhanced photon blockade}, and etc.
In recent years, the influence of $\mathcal{P}\mathcal{T}$ symmetry on the quantum properties such as quantum enhanced sensing \cite{PT symmetry magnetometry}, magnon induced transparency and amplification \cite{magnon induced transparency and amplification} as well as magnon blockade \cite{magnon blockade 3} in hybrid magnonic systems has attracted lots of attention. 

In the last decade, several theoretical hybrid magnonic systems have been proposed to generate nonclassical antibunching magnon statistics and magnon blockade based on the nonlinearity of the system, in which the competition between the nonlinear coupling strength and the dissipation rates is a challenge\cite{magnon blockade 1, magnon blockade 2, magnon blockade 3, magnon blockade 4, magnon blockade 5}.

In the present contribution, we propose a $\mathcal{P}\mathcal{T}$-symmetric cavity magnon polariton and show that in the $\mathcal{P}\mathcal{T}$- symmetric phase transition a perfect magnon blockade can be easily achieved under a small value of nonlinearity relaxing the requirement conditions in conventional magnon blockade. By studying the quantum statistics of photons, we show that a simultaneous photon blockade can occur with magnon blockade, which  experimentally can be considered as a criteria for magnon blockade.

The paper is organized as follows. In Sec. (\ref{sec2}), the system and Hamiltonian are described. Magnon and photon blockade in the proposed scheme are investigated in Sec. (\ref{sec3}). Then, in Sec. (\ref{sec4}) we represent the physical description of magnon and photon blockade in the $\mathcal{P}\mathcal{T}$-symmetric phase.  The experimental realization of the proposed scheme are discussed in Sec. (\ref{sec5}). Finally, the summary, conclusion, and outlooks
are mentioned in Sec. Sec. (\ref{sec6}).

\begin{figure}
\includegraphics[width=12cm]{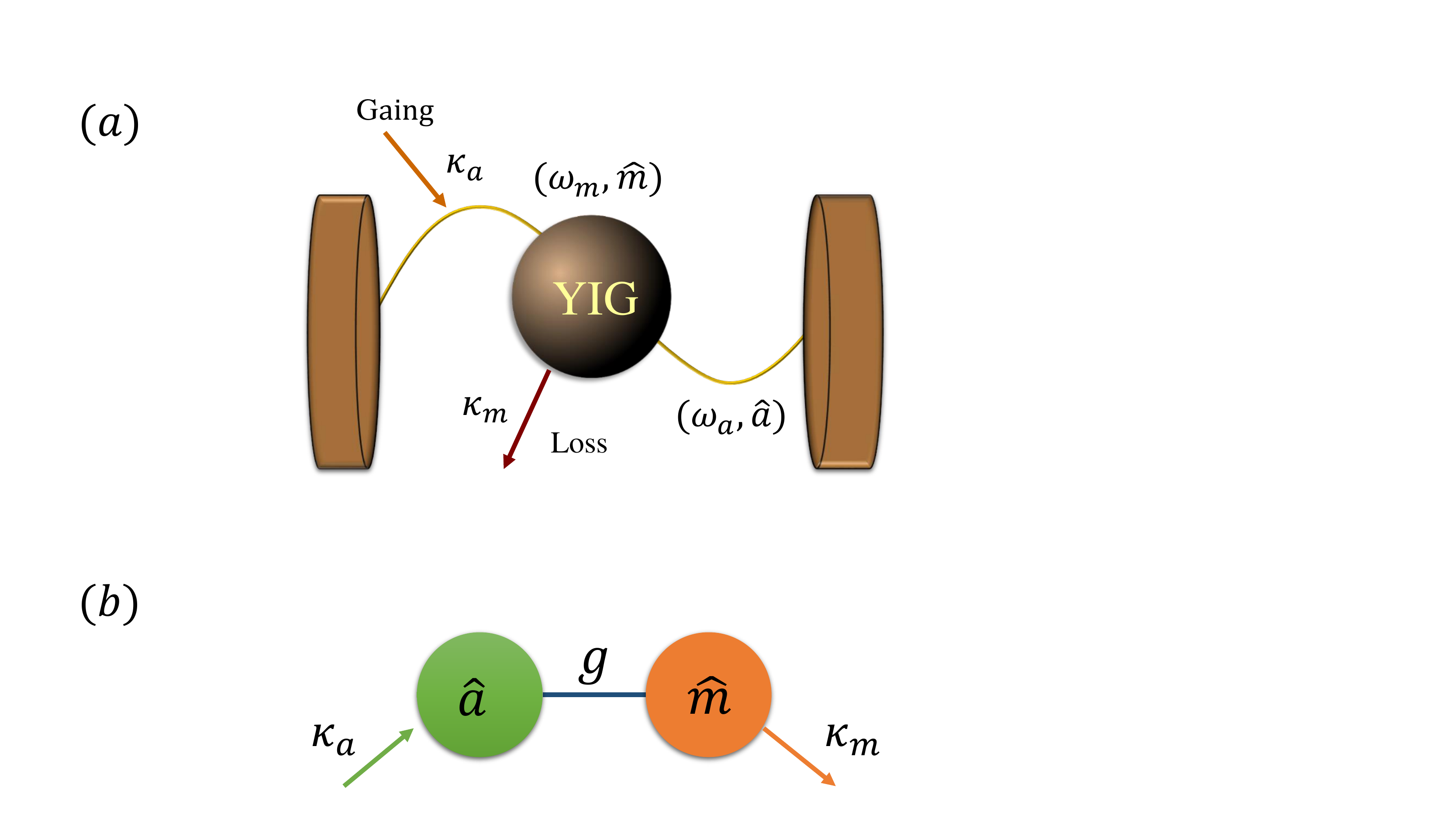}
\caption{(Color online)  (a) A schematic of a $\mathcal{P}\mathcal{T}$-symmetric cavity magnonic system, in which by changing the position of the YIG sphere the magnon-photon coupling strength can be controlled. (b) A Diagram of coupled cavity-magnon mode, in which the cavity mode is amplified with $\kappa_a$ while the magnon mode is dissipated with $\kappa_m$. }
\label{Fig1}
\end{figure}

\section{The system and Hamiltonian}\label{sec2}
Fig. (\ref{Fig1}) shows a schematic of the proposed cavity magnonic system, in which a single YIG sphere is coupled to a microwave cavity. The Hamiltonian of such a system is given by
\begin{eqnarray}\label{H}
\hat{H}=\hat{H}_a+\hat{H}_m+\hat{H}_{int},
\end{eqnarray}
where $\hat{H}_a$ is free energy of the cavity mode and assuming that the cavity is optimized to support a single photon mode is described by $\hat{H}_a=\hbar \omega_a\hat{a}^{\dagger}\hat{a}$. $\omega_a$ being the cavity mode frequency and $\hat{a}(\hat{a}^{\dagger})$ represents the photon annihilation (creation) operator. In order to determine the energy of the YIG  sphere , the Zeeman energy, demagnetization energy, as well as magnetocrystallian anisotrpy energy have to be included. Consequently, the Hamiltonian of the YIG sphere is described by \cite{theory of magnon kerr effect}
\begin{eqnarray}\label{Hm1}
H_m=-\int_{V_m} \textbf{M}.\textbf{B}_0 d \textbf{V}-\frac{\mu_0}{2}\int_{V_m} \textbf{M}.(\textbf{H}_{de}+\textbf{H}_{an}) d\textbf{V},
\end{eqnarray}
where $V_m$ is the volume of the YIG sphere, $\textbf{M}$ is the sample magnetization, $\textbf{B}_0$ is the static magnetic field applied in the $z$ direction, and $\mu_0$ is the magnetic permeability of free space.
The first term elucidates the Zeeman energy, while the second part represents demagnetization and magneto crystallian anisotropy energies with $\textbf{H}_{de}$ and $\textbf{H}_{an}$ being the demagnetization field induced by the static magnetic field and the anisotropic field caused by the magnetocrystallian anisotropy in YIG. In the proposed scheme, the YIG  sphere is placed inside a microwave cavity with uniform external bias magnetic field , which is applied to the YIG sphere to produce a homogeneous magnonic mode, the so-called Kittel mode. For such a homogeneous magnonic mode, by using the Holstein-Primakoff representation the Hamiltonian can be written in terms of the magnon annihilation and creation operators $\hat{m}$ and $\hat{m}^{\dagger}$ with frequency $\omega_m$ as \cite{theory of magnon kerr effect}
\begin{eqnarray}\label{Hm2}
\hat{H}_m=\hbar \omega_m \hat{m}^{\dagger}\hat{m}+\hbar \chi (\hat{m}^{\dagger}\hat{m})^2.    
\end{eqnarray}
The second term describes the magnon Kerr effect originating from the magnetocrystallian anisotropy  in the YIG sphere. The coefficient $\chi$ characterizing the strength of the nonlinear magnon effect is given by $\chi=\frac{\mu_0 \chi _{an}\gamma^2}{M^2V_m}$, where $\chi_{an}$ is the first order anisotropy constant of YIG, $\gamma=g_e \mu_B$ is the gyromagnetic ratio with $g_e$ being the Lande factor and $\mu_B$ denoting the Bohr magneton.
Finally, the magnon-cavity mode interaction Hamiltonian in the second quantization can be written as $\hat{H}=\hbar g (\hat{a}^{\dagger}\hat{m}+\hat{a}\hat{m}^{\dagger})$, in which by assuming $g<< \omega_a, \omega_m$, is simplified under the rotating wave approximation (RWA) \cite{hybrid magnonic system 2}. The strength of magnon-photon coupling is given by $g=\frac{\gamma B_0}{2}\sqrt{2sN}$ where $s$ and $N$ respectively stand for the spin angular momentum $s=5/2$ on each unit cell of the magnetic material and the number of unit cells with spin $s$ in the YIG sphere \cite{magnon meets the superconducting qubit}. It worth emphasizing that in contrast to the Kerr nonlinear coefficient, which is inversely proportional to the YIG volum $(\chi \propto V_M^{-1})$, the coupling strength is proportional to the square root of the YIG volume $(g \propto \sqrt{N}\propto \sqrt{V_m})$. Consequently, reducing the size of the YIG sphere leads to the Kerr nonlinear coefficient increment and coupling strength decrement.
Finally, it is supposed that the two magnonic and photonic modes in the proposed scheme are dissipated with $\kappa_a$ and $\kappa_m$  ($\kappa_a, \kappa_m>0$ ), respectively. By including the dissipation processes, the system can be effectively described by the following non-Hermitian Hamiltonian
\begin{eqnarray}\label{final H}
 \hat{H}&=&\hbar (\omega_a-i\kappa_a)\hat{a}^{\dagger}\hat{a}+\hbar(\omega_m-i\kappa_m)\hat{m}^{\dagger}\hat{m}+\hbar \chi (\hat{m}^{\dagger}\hat{m})^2\nonumber\\
 &&+\hbar g (\hat{a}^{\dagger}\hat{m}+\hat{a}\hat{m}^{\dagger})  +\hbar \Omega (\hat{m}^{\dagger}e^{-i\omega_dt}+\hat{m}e^{i\omega_dt}),
\end{eqnarray}
where it has been supposed that the magnonic mode is driven with an external magnetic field \cite{theory of magnon kerr effect}. The last term describes such an interaction in which the coupling strength of the magnon drive field (whose frequency and magnitude are respectively given by $\omega_d$ and $B_d$) with the magnonic mode of the YIG sphere is $\Omega= \frac{\eta}{2}B_d$, where $\eta=\frac{\gamma}{2}\sqrt{5N}$.

In the following, inspired by several studies based on $\mathcal{P}\mathcal{T}$-symmetric devices \cite{pt symmetry app 1, low power optical isolation 1, low power optical isolation 2, single mode microcavity laser 1, single mode microcavity laser 2, loss-induced or gain induced transparency 1, loss-induced or gain induced transparency 2, PT symmetry enhanced photon blockade, magnon induced transparency and amplification}, we suppose that the photonic mode is amplified such that there is no loss for the cavity mode and there is gain instead (Fig\ref{Fig1}). In this case, the photonic mode dissipation rate is negative ($\kappa_a<0$), while in order to provide the $\mathcal{P}\mathcal{T}$-symmetric conditions in the system, the magnonic mode dissipation rate is still positive ($\kappa_m>0$). By the substitution of $\kappa_m=-\kappa_a$ in Eq. (\ref{final H}), it can be found that this Hamiltonian under the condition of a very weak nonlinearity, $\chi<<1$, and switching off the driving field, $\Omega_d=0$, satisfies the $\mathcal{P}\mathcal{T}$-symmetric.
The Hamiltonian with positive dissipation rates $(\kappa_a,\kappa_m>0)$ under the above mentioned conditions can be written as
\begin{eqnarray}\label{Hmatrix}
 \hat{H} = \begin{pmatrix}
 \hat{a}^{\dagger} && \hat{m}^{\dagger}
 \end{pmatrix}  
 \begin{pmatrix}
     \omega_a-i\kappa_a && g\\
     g && \omega_m-i\kappa_m
 \end{pmatrix}
 \begin{pmatrix}
     \hat{a}\\
     \hat{m}
 \end{pmatrix},
\end{eqnarray}
which its eigenvalues are
\begin{eqnarray}\label{eigenvalues}
    \omega_{1,2}&=&\frac{1}{2}\bigg(\omega_a+\omega_m-i(\kappa_a+\kappa_m)\bigg)\nonumber\\
    &&\pm \frac{1}{2}\sqrt{4g^2-\bigg[i(\omega_a-\omega_m)+(\kappa_a-\kappa_m)\bigg]^2}.
\end{eqnarray}

\begin{figure}
\includegraphics[width=8cm]{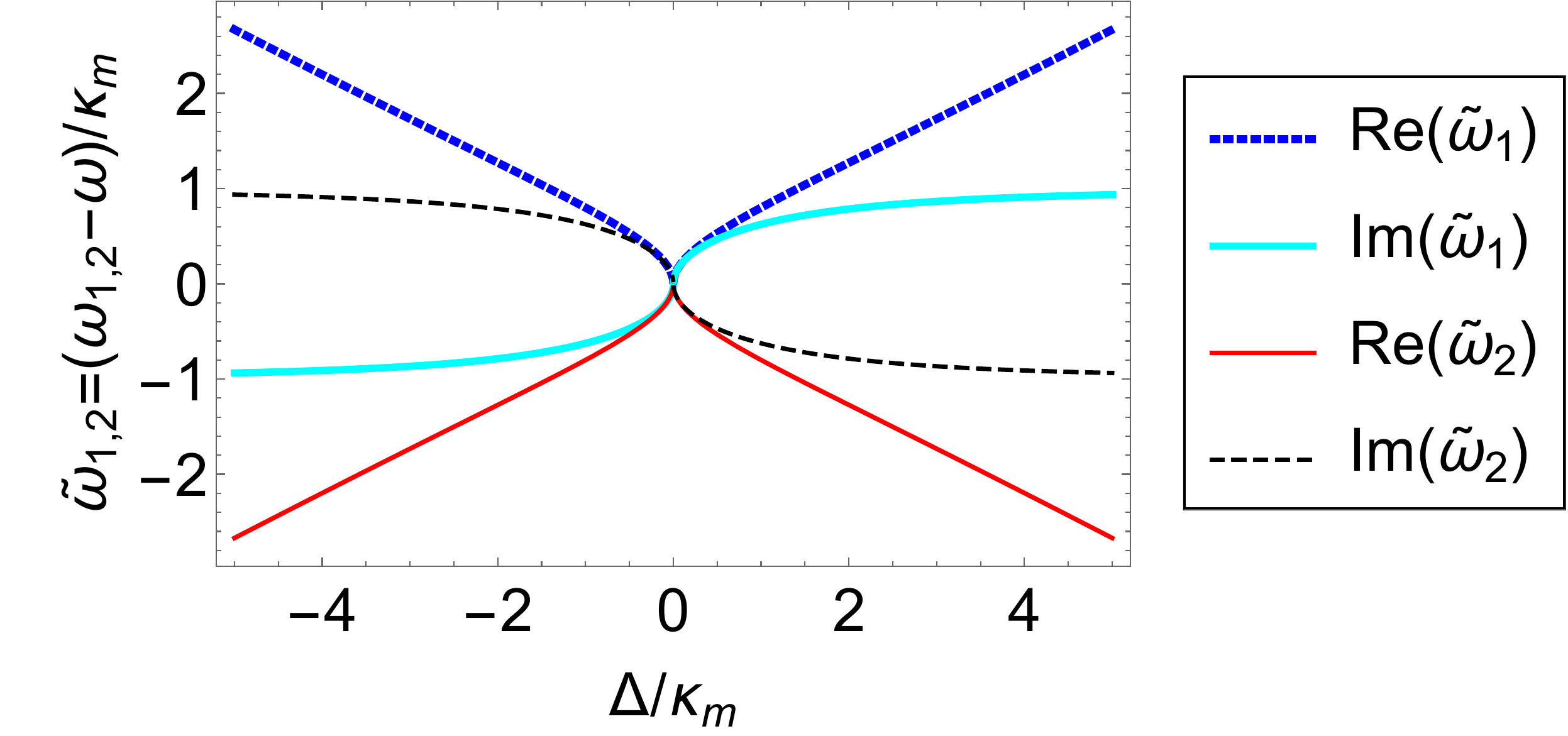}
\caption{(Color online) The real and imaginary part of $\tilde{\omega}_{1,2}=(\omega_{1,2}-\omega)/\kappa_m$ versus the normalized detuning, $\Delta/\kappa_m$. The selected parameters are as follows: $\kappa_a/\kappa_m=-1$, $g/\kappa_m=1$. The blue-dotted, cyan solid line, red solid line, and black dotted respectively correspond to $Re(\tilde{\omega}_1)$, $Im(\tilde{\omega}_1)$, $Re(\tilde{\omega}_2)$, and $Im(\tilde{\omega}_2)$.}
\label{Fig2}
\end{figure}

\begin{figure}
\includegraphics[width=8cm]{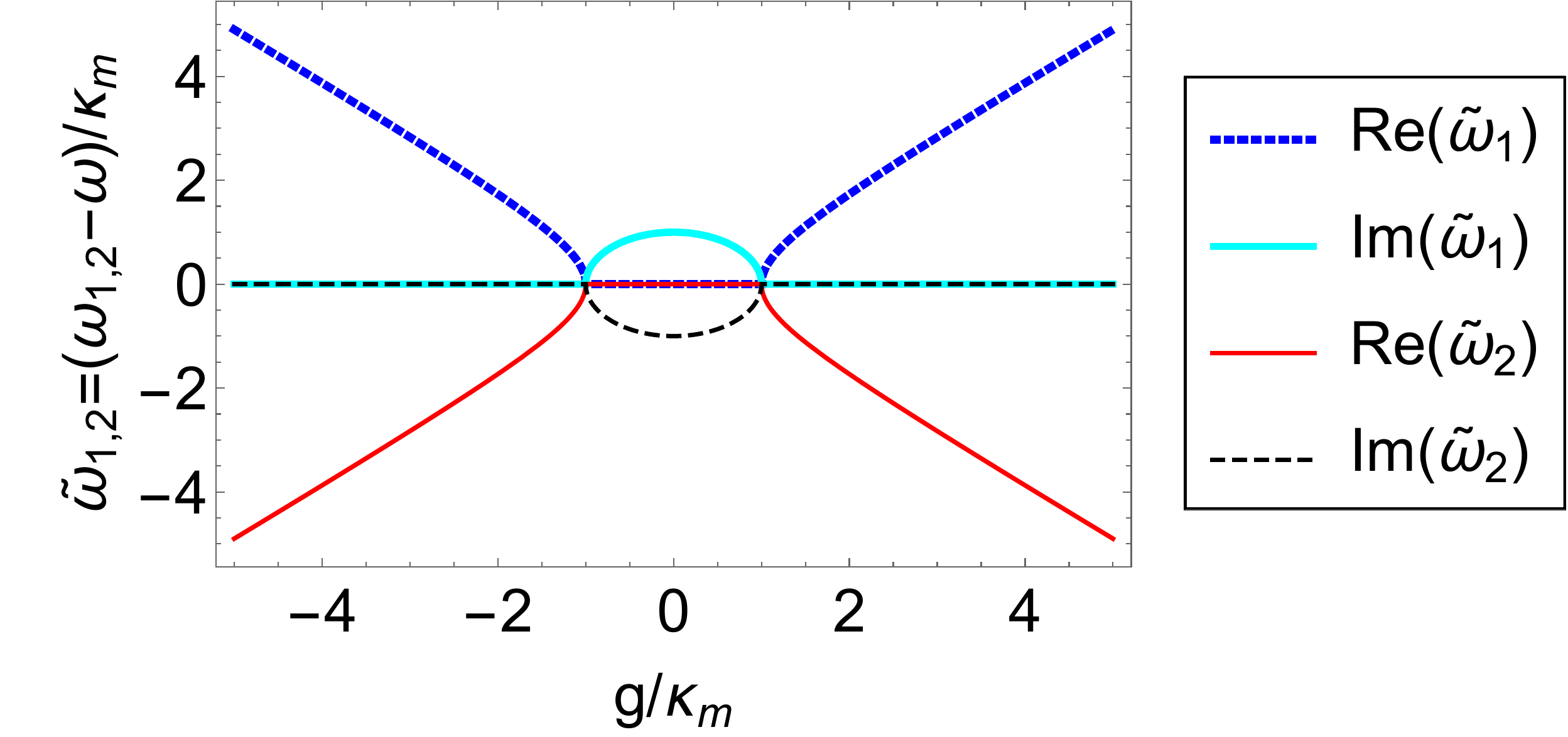}
\caption{(Color online) The real and imaginary part of $\tilde{\omega}_{1,2}=(\omega_{1,2}-\omega)/\kappa_m$ versus the normalized coupling strength, $g/\kappa_m$. The selected parameters are as follows: $\kappa_a/\kappa_m=-1$, $\Delta/\kappa_m=0$. The blue-dotted, cyan solid line, red solid line, and black dotted respectively correspond to $Re(\tilde{\omega}_1)$, $Im(\tilde{\omega}_1)$, $Re(\tilde{\omega}_2)$, and $Im(\tilde{\omega}_2)$.}
\label{Fig3}
\end{figure}

\begin{figure}
\includegraphics[width=8cm]{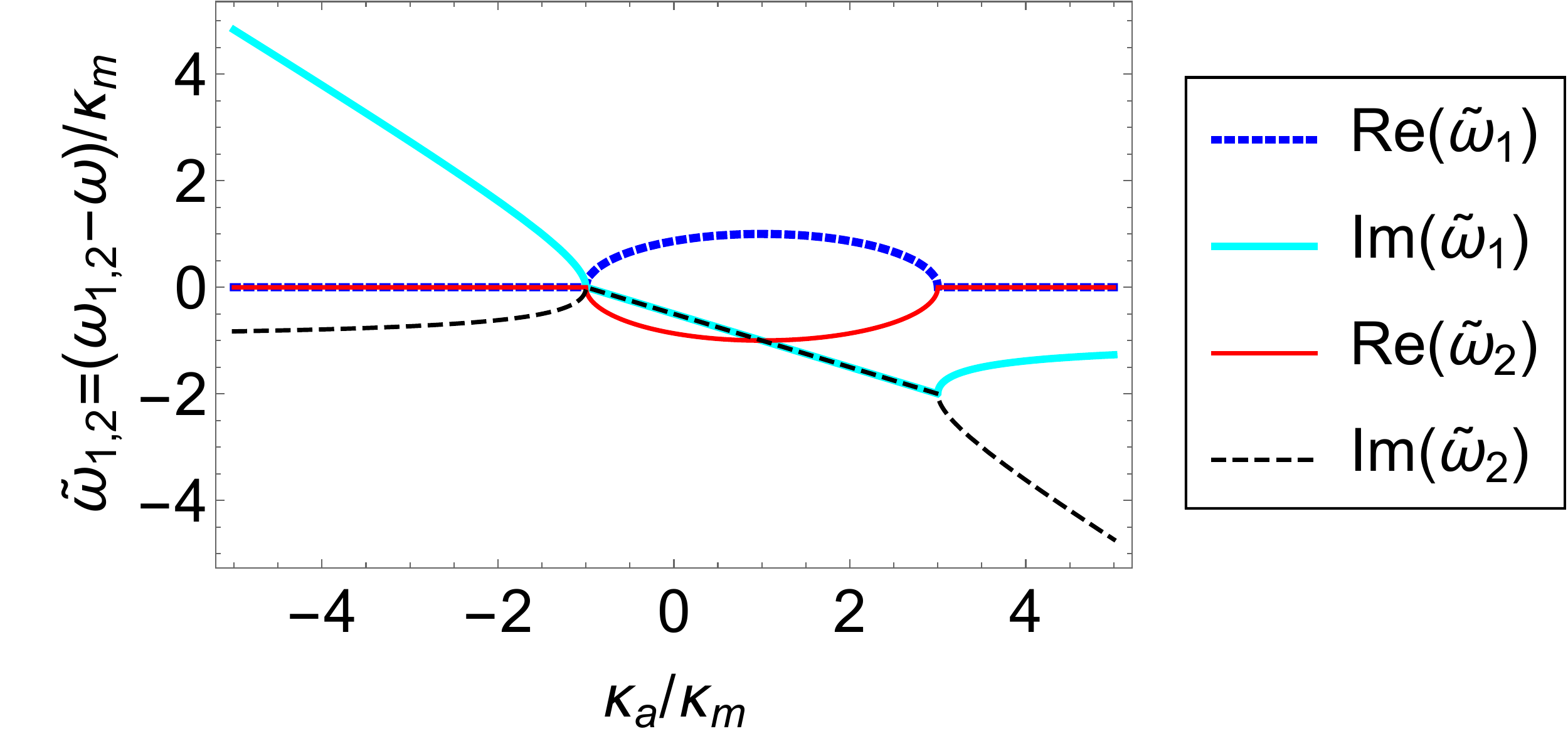}
\caption{(Color online) The real and imaginary part of $\tilde{\omega}_{1,2}=(\omega_{1,2}-\omega)/\kappa_m$ versus the normalized cavity mode dissipation rate, $\kappa_a/\kappa_m$. The selected parameters are as follows: $\Delta/\kappa_m=0$, $g/\kappa_m=1$. The blue-dotted, cyan solid line, red solid line, and black dotted respectively correspond to $Re(\tilde{\omega}_1)$, $Im(\tilde{\omega}_1)$, $Re(\tilde{\omega}_2)$, and $Im(\tilde{\omega}_2)$.}
\label{Fig4}
\end{figure}

In Eq. (\ref{eigenvalues}), the two eigenfrequencies, $\omega_{1,2}$, depend on $\omega_a$, $\omega_m$, $\kappa_a$, $\kappa_m$, and $g$. This equation evidently shows that to obtain purely real eigenvalue spectra, it is required that $\omega_a=\omega_m\equiv \omega$, $\kappa_a=-\kappa_m$ and $g>\kappa_a$ which can be well stablished in the system \cite{observation of pt-magnonic}. Achieving $\omega_a=\omega_m$ is experimentally possible and in such a case that the magnon frequency is close to the photon frequency, the strong interaction between magnon and cavity photon mixes each degree of freedom and creates the hybridized states of the cavity magnon-polariton \cite{observation of pt-magnonic}. The second condition, $\kappa_a=-\kappa_m$,  characterizes the balanced gain and loss for the two coupled modes, cavity photon and magnon. In this case, the eigenfrequencies reduces to $\tilde{\omega}_{1,2}=\omega_a\pm \sqrt{4g^2-\kappa_a^2}$. Consequently, three different scenarios can be studied: (i) if $g=\kappa_a$, the two frequencies collides to the central frequency $\omega_a$, (ii) if $g>\kappa_a$, the system is in the unbroken $\mathcal{P}\mathcal{T}$-symmetric phase, and finally (iii) if $g<\kappa_a$ the eigenfrequencies becom complex and the $\mathcal{P}\mathcal{T}$-symmetric is spontaneously broken. Consequently, $g=\kappa_a$ often referring to the spontaneous $\mathcal{P}\mathcal{T}$ symmetric point, is called exceptional point (EP). In order to obtain more insight, the real and imaginary part of ${\omega}^{\prime}_{1,2}=(\omega_{1,2}-\omega)/\kappa_m$ versus the normalized detuning, $\Delta/\kappa_m$, normalized cavity-magnon coupling strength, $g/\kappa_m$, and the normalized cavity mode dissipation rate, $\kappa_a/\kappa_m$, are respectively represented in Figs. (\ref{Fig2},\ref{Fig3},\ref{Fig4}), which all show that under the above mentioned conditions the real and imaginary parts of the eigenvalues coincide.

It should be mention that $g=\kappa_a$ is not the only EP in such a system. Another selection as $g=(\kappa_a-\kappa_m)$, which leads to the eigenfrequencies coincidence can also known as the exceptional point, when $\omega_a=\omega_m$, and $\kappa_a\neq -\kappa_m$. Evidently, the threshold of the symmetry breaking strongly depends on the cavity and magnon mode dissipation rates. As in the last decade several studies such as $\mathcal{P}\mathcal{T}$-symmetric enhanced sensing \cite{PT symmetry magnetometry}, photon blockade \cite{PT symmetry enhanced photon blockade}, and etc have been studied in systems with $\mathcal{P}\mathcal{T}$-symmetric properties,  the $\mathcal{P}\mathcal{T}$-symmetric phase and EP influence the dynamics of the system. In the following, we study the behaviour of magnon as well as photon blockade on the EP in the proposed scheme.

\section{magnon and photon blockade in $\mathcal{P}\mathcal{T}$-symmetric cavity magnonic system}\label{sec3}

By rotating the Hamiltonian, Eq. (\ref{final H}), in the frame with the frequency of the magnonic mode driving field, $\omega_d$, we obtain 
\begin{eqnarray}\label{rotated H}
    \hat{H}_{rot}&=&\hbar (\Delta_a-i\kappa_a)\hat{a}^{\dagger}\hat{a}+\hbar(\Delta_m-i\kappa_m)\hat{m}^{\dagger}\hat{m}+\hbar \chi(\hat{m}^{\dagger}\hat{m})^2\nonumber\\
    &&+\hbar g(\hat{a}^{\dagger}\hat{m}+\hat{a}\hat{m}^{\dagger})+\hbar \Omega_d (\hat{m}^{\dagger}+\hat{m}),
\end{eqnarray}
in which $\Delta_{a(m)}=\omega_{a(m)}-\omega_d$. Evidently, in the cavity magnon polariton with $\omega_a=\omega_m$ we obtain the same frequency detunings $\Delta_a=\Delta_m\equiv \Delta$.

The magnon (photon) statistics and consequently magnon (photon) blockade effect is characterized by the equal-time second order correlation function
\begin{eqnarray}
    g^{(2)}_m(0)=\frac{\langle \hat{m}^{\dagger}\hat{m}^{\dagger}\hat{m}\hat{m}\rangle}{\langle \hat{m}^{\dagger}\hat{m}\rangle ^2}=\frac{\sum_m m(m-1)P_m}{\bigg(\sum_m mP_m\bigg)^2},
\end{eqnarray}

\begin{eqnarray}
    g^{(2)}_a(0)=\frac{\langle \hat{a}^{\dagger}\hat{a}^{\dagger}\hat{a}\hat{a}\rangle}{\langle \hat{a}^{\dagger}\hat{a}\rangle ^2}=\frac{\sum_n n(n-1)P_n}{\bigg(\sum_n nP_n\bigg)^2}.
\end{eqnarray}

The equal time second order correlation functions, $g^{(2)}_{m(a)}(0)$ , represents the quantum statistics of magnon (photon) excitations such that $g^{(2)}_{m(a)}(0)>1$ and $g^{(2)}_{m(a)}(0)=1$ respectively correspond to the super-Poissonian and Poissonian statistics. While for a sub-Poissonian  statistics as  a non-classical feature  $g^{(2)}_{m(a)}(0)<1$ should be provided. $g^{(2)}_{m(a)}(0)<1$ expresses the unwilling of simultaneous magnon (photon) excitation and hence $g^{(2)}_{m(a)}(0)\to 0$ is referred to a perfect magnon (photon) blockade. 
The dynamics of the system can be determined by numerically solving the master equation $\frac{\partial}{\partial t}\rho=[\rho,\tilde{H}_{rot}]+\mathcal{L}\rho$, in which $\rho$ is the system density matrix , $\tilde{H}_{rot}$ is given by
\begin{eqnarray}\label{HRT}
    \tilde{H}_{rot}&=&\hbar \Delta_a\hat{a}^{\dagger}\hat{a}+\hbar \Delta_m\hat{m}^{\dagger}\hat{m}++\hbar \chi(\hat{m}^{\dagger}\hat{m})^2\nonumber\\
    &&+\hbar g (\hat{a}^{\dagger}\hat{m}+\hat{a}\hat{m}^{\dagger})+\hbar \Omega_d(\hat{m}^{\dagger}+\hat{m}).
\end{eqnarray}
In this approach, the dissipation rates are not included in the Hamiltonian but are described by the Liouvillian superoperators as 
$\hat{\mathcal{L}}{\rho}=\kappa_a(2\hat{a}\hat{\rho}\hat{a}^{\dagger}-\hat{a}^{\dagger}\hat{a}\hat{\rho}-\hat{\rho}\hat{a}^{\dagger}\hat{a})+ \kappa_m(2\hat{m}\hat{\rho}\hat{m}^{\dagger}-\hat{m}^{\dagger}\hat{m}\hat{\rho}-\hat{\rho}\hat{m}^{\dagger}\hat{m})$, in which as is considered through this study, the thermal photon and magnon excitation  are zero, $n_{th}^{(a)}=n_{th}^{(m)}=0$.

Another approach is analytically solving the Schrodinger equation $i\hbar \frac{\partial}{\partial t}\vert \psi \rangle =\hat{H}_{rot}\vert \psi \rangle$. To analytically determine the magnon (photon) statistics as the main criteria of magnon (photon) blockade, we consider the weak driving limit $\Omega_d <<\chi, g, \Delta$, in which the state of the system can be truncated in a few excitation subspace  as
\begin{eqnarray}\label{psi}
    \vert \psi \rangle&=& C_{00}\vert 0 0 \rangle +C_{10}\vert 1 0  \rangle + C_{01}\vert 0 1  \rangle + C_{11}\vert 1 1  \rangle + C_{20}\vert 2 0  \rangle\nonumber\\
    && + C_{02}\vert  0 2  \rangle  + C_{12}\vert 1 2  \rangle+ C_{21}\vert 2 1  \rangle +C_{22}\vert 2 2  \rangle,
\end{eqnarray}
with $C_{mn}$ representing the probability amplitude of $m$ magnon and $n $ photon number excitations in the system. The probability amplitudes $C_{mn}$ in the state $\vert \psi \rangle $ can be obtained by solving the Schrodinger equation $i\hbar \frac{\partial}{\partial t} \vert \psi\rangle =\hat{H}_{rot}\vert \psi \rangle$. To obtain the probability amplitudes in the steady state, we substitute the truncated state , Eq. (\ref{psi}), into the Schrodinger equation. Consequently, in the steady state we obtain
\begin{widetext}
\begin{eqnarray}
\dot{C}_{10}&=&C_{10}(\Delta_m-i\kappa_m)+\chi C_{10}+g C_{01}+\Omega\sqrt{2} C_{20}+\Omega=0,\\
\nonumber\\%
\dot{C}_{01}&=&C_{01}(\Delta_a-i\kappa_a)+g C_{10}+\Omega C_{11}=0,\\
\nonumber\\%
\dot{C}_{11}&=&C_{11}(\Delta_a-i\kappa_a)+C_{11}(\Delta_m-i\kappa_m)+\chi C_{11}+\sqrt{2}g(C_{20}+C{02})+\Omega\sqrt{2}C_{21}+\Omega C_{01}=0,\\
\nonumber\\%
\dot{C}_{20}&=&2C_{20}(\Delta_m-i\kappa_m)+4\chi C_{20}+\sqrt{2}g C_{11} +\sqrt{2}\Omega C_{10}=0,\\ 
\nonumber\\%
\dot{C}_{02}&=&2C_{02}(\Delta_a-i\kappa_a)+g\sqrt{2}C_{11}+\Omega C_{12}=0,\\
\nonumber\\%
\dot{C}_{12}&=&C_{12}(\Delta_m-i\kappa_m)+2C_{12}(\Delta_a-i\kappa_a)+\chi C_{12}+2g C_{21}+\Omega C_{22}+\Omega C_{02}=0,\\
\nonumber\\%
\dot{C}_{21}&=&2C_{21}(\Delta_m-i\kappa_m)+C_{21}(\Delta_a-i\kappa_a))+4\chi C_{21}+2g C_{12}+\sqrt{2}\Omega C_{11}=0,\\
\nonumber\\%
\dot{C}_{22}&=&2C_{22}(\Delta_a-i\kappa_a)+2C_{22}(\Delta_m-i\kappa_m)+4\chi C_{22}+\Omega \sqrt{2} C_{12}=0.
\end{eqnarray}
\end{widetext}

By considering the truncated state, Eq. (\ref{psi}), under the assumption of weak driving limit, the equal-time second order correlation functions, $g_m^{(2)}(0)$ and $g_a^{(2)}(0)$ can be reduced as follows
\begin{eqnarray}
    g_m^{(2)}(0)&=&\frac{\sum_m m(m-1)P_m}{\bigg(\sum_m mP_m\bigg)^2}\\
    &\simeq & \frac{2\vert C_{20}+C_{21}+C_{22}\vert^2}{\bigg(\vert C_{10}+C_{11}\vert^2+2\vert C_{20}+C_{21}+C_{22}\vert^2\bigg)^2},\nonumber
\end{eqnarray}

\begin{eqnarray}
    g_a^{(2)}(0)&=&\frac{\sum_n n(n-1)P_n}{\bigg(\sum_n nP_n\bigg)^2}\\
    &\simeq & \frac{2\vert C_{02}+C_{12}+C_{22}\vert^2}{\bigg(\vert C_{01}+C_{11}\vert^2+2\vert C_{02}+C_{12}+C_{22}\vert^2\bigg)^2}.\nonumber
\end{eqnarray}

\begin{figure}
\includegraphics[width=8cm]{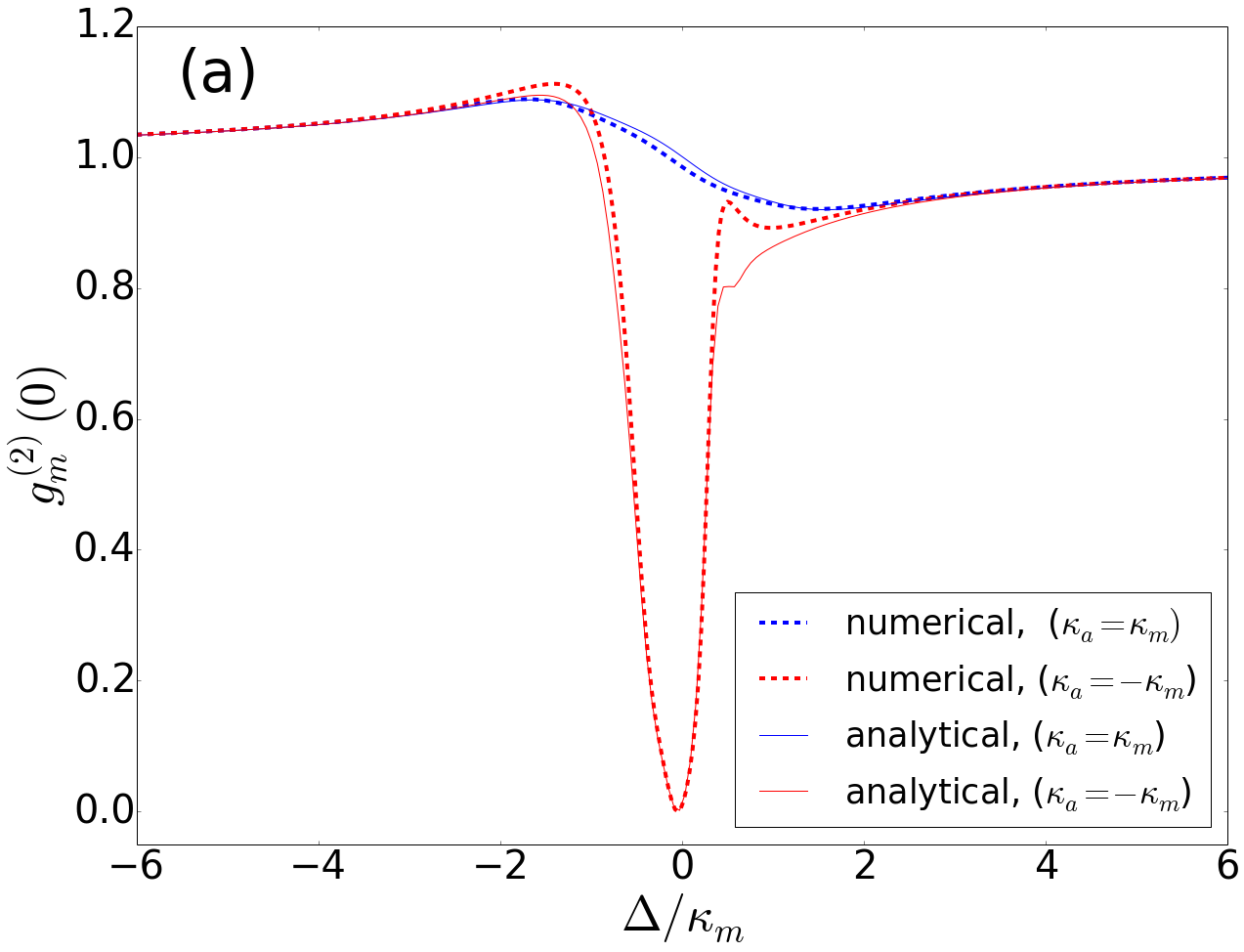}
\hspace{20mm}
\includegraphics[width=8cm]{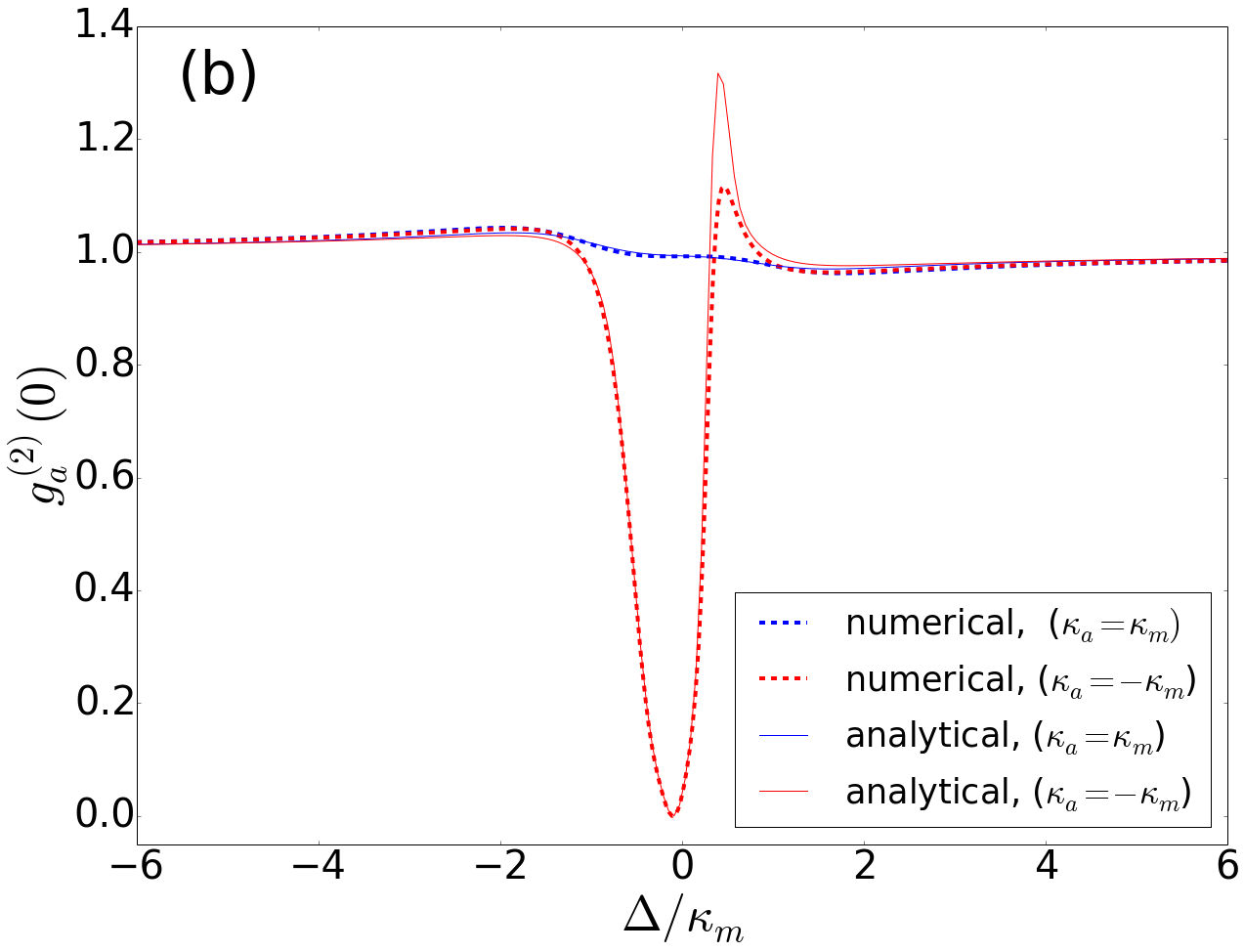}
\caption{(Color online) Equal-time second-order correlation function of (a) magnons and (b) photons, $g_m^{(2)}(0)$ and $g_a^{(2)}(0)$, versus the normalized detuning, $\Delta/\kappa_m$. The
blue and red lines respectively refer to $\kappa_a=\kappa_m$ and $\kappa_a=-\kappa_m$. The
other selected parameters are as follows: $g/\kappa_m=1$, $\chi/\kappa_m=0.1$, and $\Omega_d/\kappa_m=0.01$.
 The solid and dotted lines respectively represent the analytical and 
numerical results.}
\label{Fig58}
\end{figure}

\begin{figure}
\includegraphics[width=8cm]{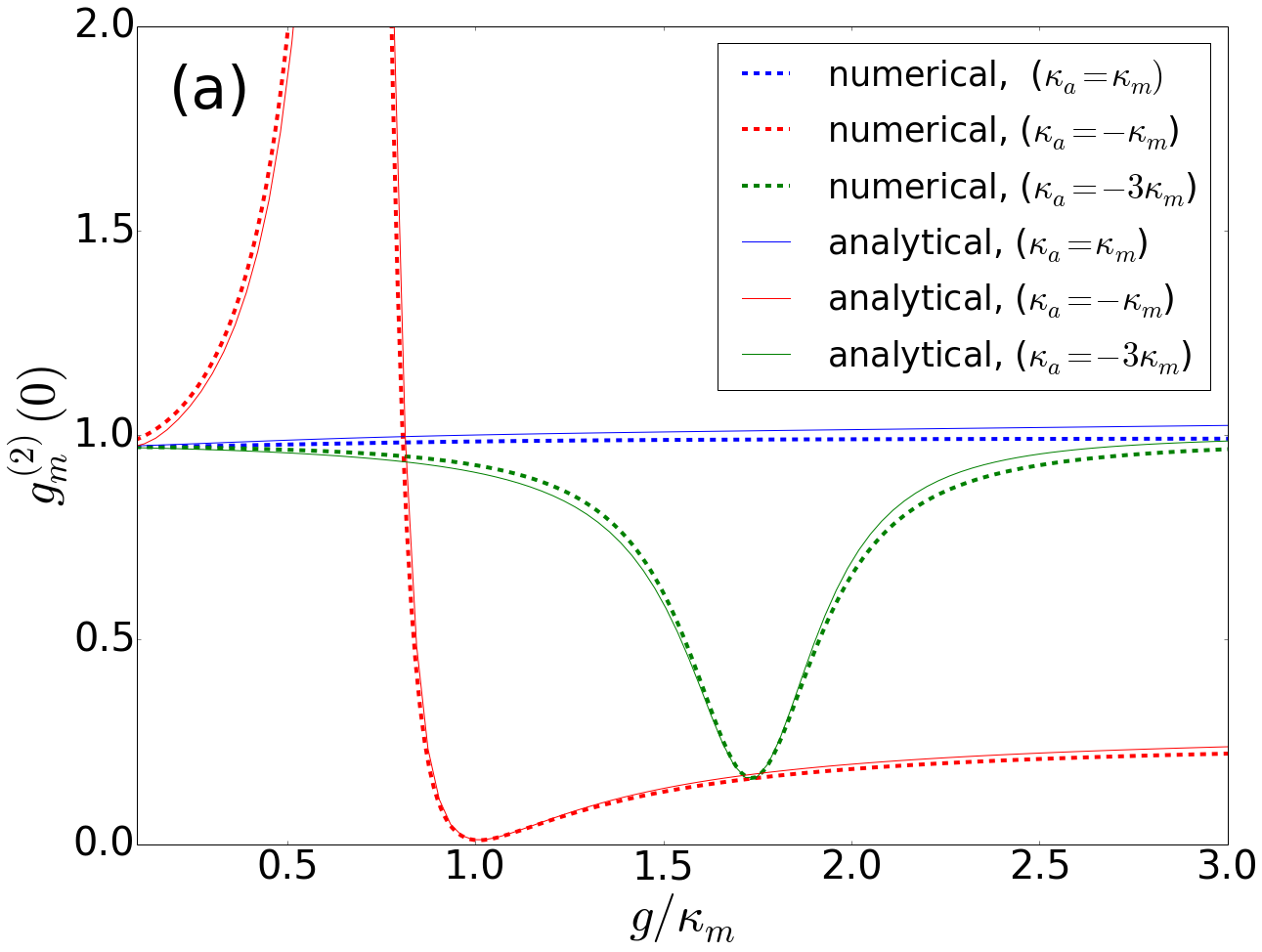}
\hspace{10mm}
\includegraphics[width=8cm]{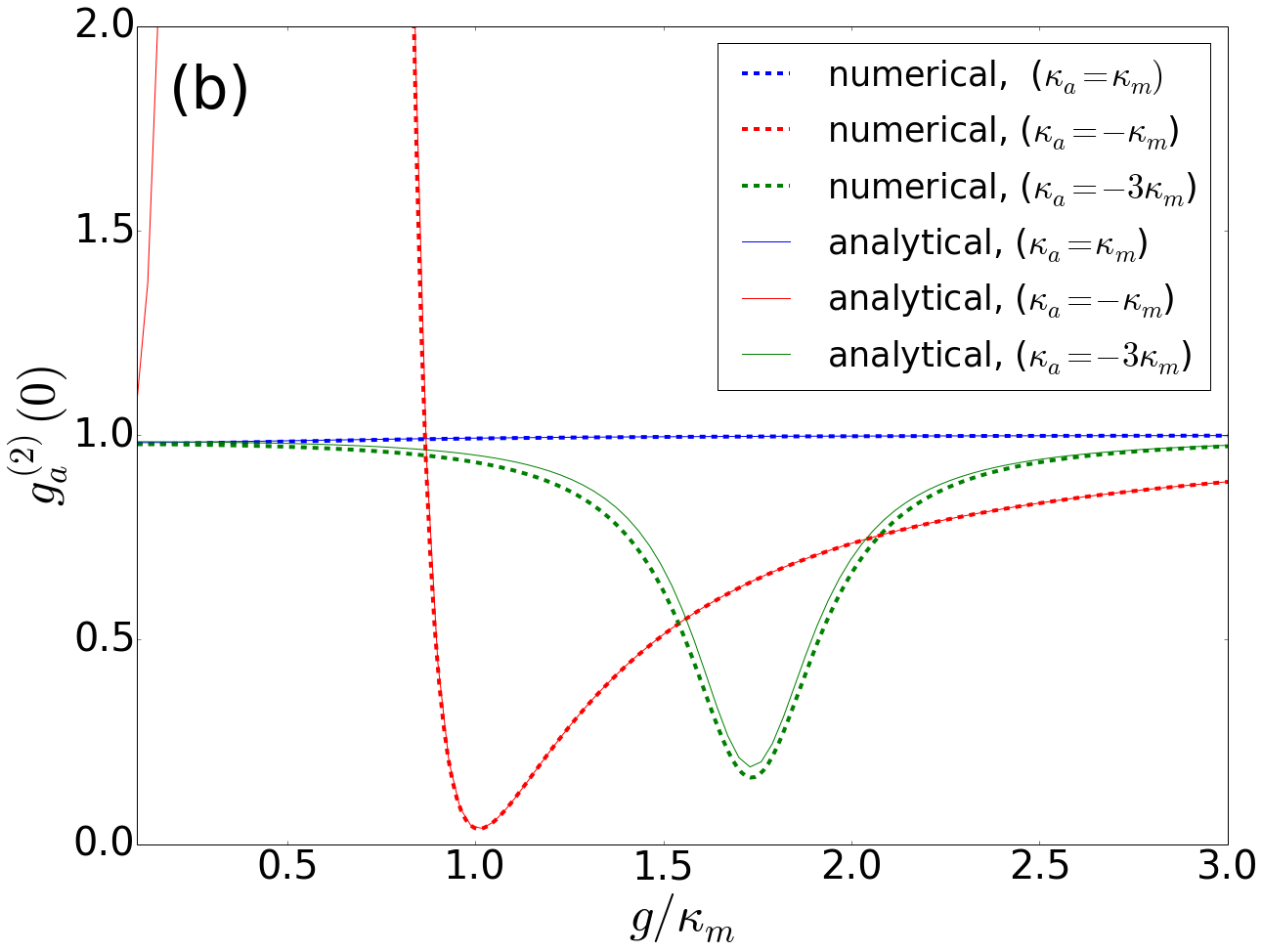}
\caption{(Color online) Equal-time second-order correlation function of (a) magnons and (b) photons, $g_m^{(2)}(0)$ and $g_a^{(2)}(0)$, versus the normalized cavity-magnon coupling strength, $g/\kappa_m$. The
blue, red and green lines respectively refer to $\kappa_a=\kappa_m$,  $\kappa_a=-\kappa_m$, and $\kappa_a=-3\kappa_m$. The
other selected parameters are as follows: $\Delta/\kappa_m=0$, $\chi/\kappa_m=0.1$, and $\Omega_d/\kappa_m=0.01$.
 The solid and dotted lines respectively represent the analytical and 
numerical results.}
\label{Fig69}
\end{figure}

\begin{figure}
\includegraphics[width=8cm]{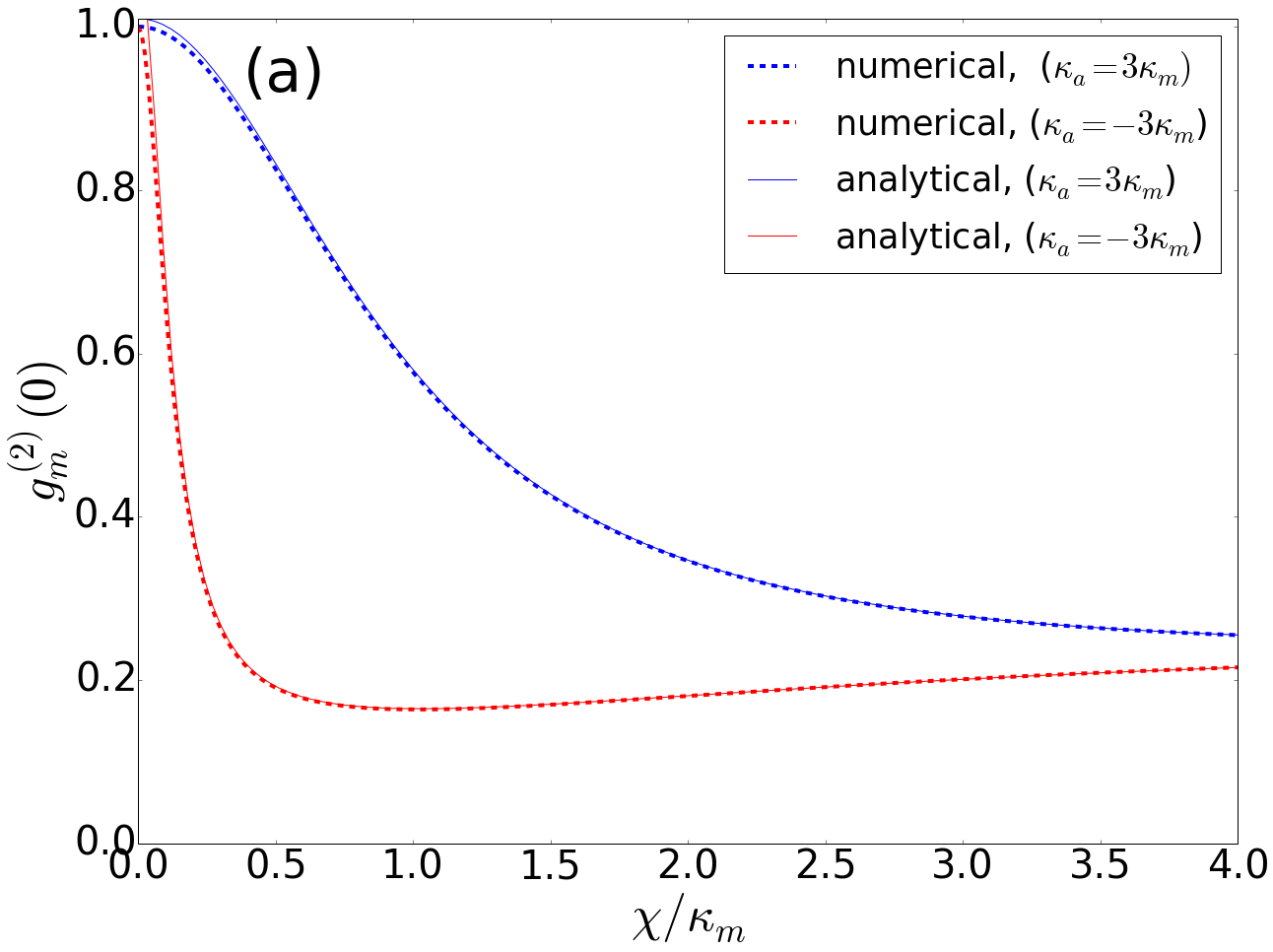}
\hspace{10mm}
\includegraphics[width=8cm]{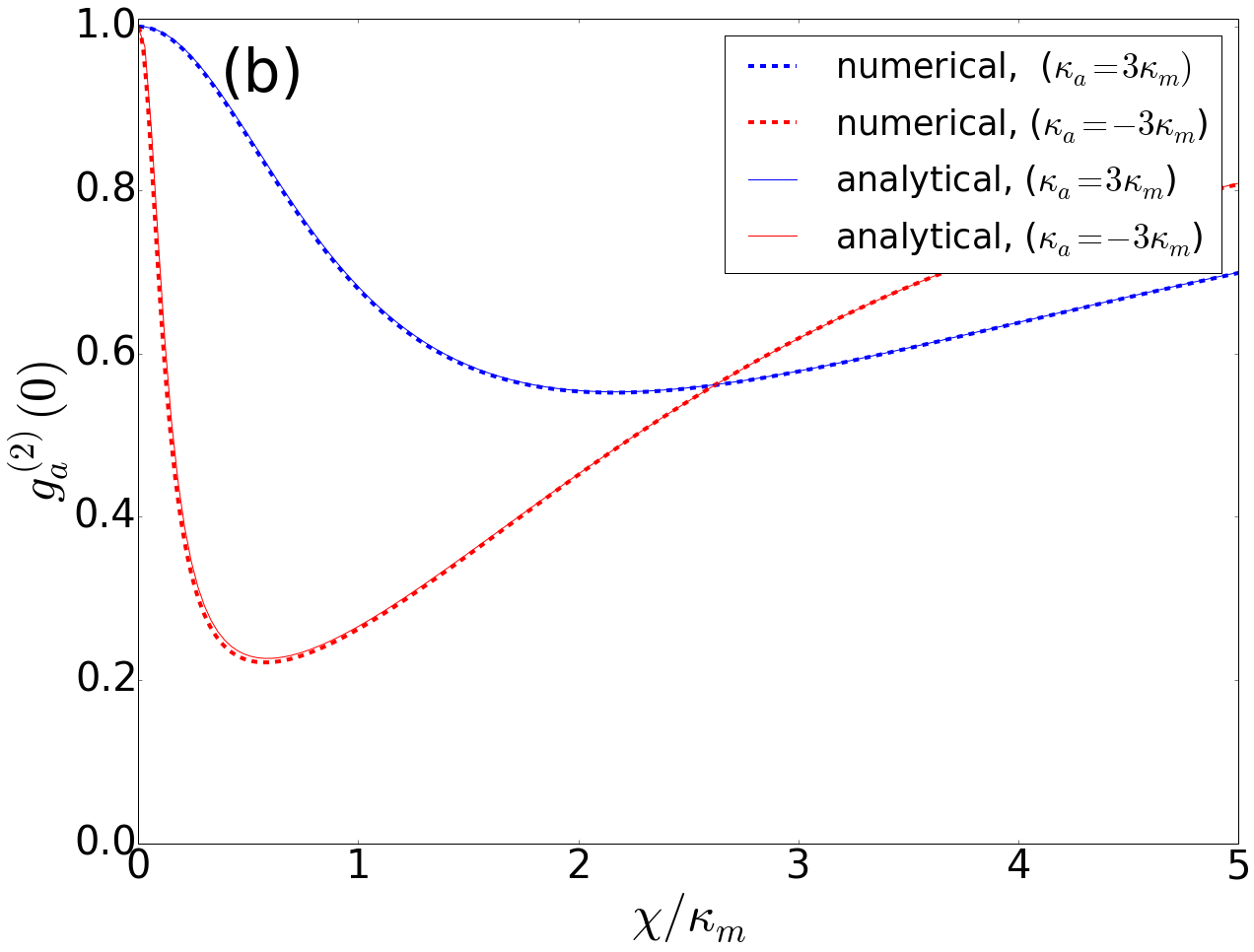}
\caption{(Color online) Equal-time second-order correlation function of (a) magnons and (b) photons, $g_m^{(2)}(0)$ and $g_a^{(2)}(0)$, versus the normalized Kerr nonlinearity, $\chi/\kappa_m$. The
blue and red lines respectively refer to $\kappa_a=3\kappa_m$ and $\kappa_a=-3\kappa_m$. The
other selected parameters are as follows: $g/\kappa_m=2$, $\Delta/\kappa_m=0$, and $\Omega_d/\kappa_m=0.01$.
 The solid and dotted lines respectively represent the analytical and 
numerical results.}
\label{Fig710}
\end{figure}

The results of the analytical approach as well as the ones obtained by numerically solving the master equation using QuTiP \cite{qutip} are represented in Figs .(\ref{Fig58}), (\ref{Fig69}), (\ref{Fig710}).
In order to obtain more insights about the behaviour of the second order magnon and photon correlation functions, $g_m^{(2)}(0)$ and $g_a^{(2)}(0)$, in the proposed $\mathcal{P}\mathcal{T}$-symmetric scheme and also see the effect of the magnon detuning, $\Delta$, the second order correlation function of magnons and photons $g_m^{(2)}(0)$ and $g_a^{(2)}(0)$ versus the normalized detuning, $\Delta/\kappa_m$, are illustrated in Figs. (\ref{Fig58})(a,b). In these figures (Figs. (\ref{Fig58})(a,b)), two different configurations are considered: (i) a coupled of dissipated cavity-magnon mode, i.e., $\kappa_a=\kappa_m$ $(g=\kappa_m)$ and (ii) a coupled dissipative magnon mode and amplified cavity mode, i.e, $\kappa_a=-\kappa_m$ $(g=\kappa_m)$. As is evident from both the analytical approximated results as well as the exact numerical ones, by selecting coupled dissipative modes ($\kappa_a=\kappa_m$), the minimum value of $g_m^{(2)}(0)$ and $g_a^{(2)}(0)$ is about $0.9$ which can be achieved by tunning  $\Delta/\kappa_m\simeq  1.5$. For the $\mathcal{P}\mathcal{T}$-symmetric coupled modes. i.e., $\kappa_a=-\kappa_m$, $g=\kappa_m$, a strong magnon and photon antibunching effect can be obtained, which are considered  as perfect magnon and photon blockade because the two magnon as well as photon excitations are strongly suppressed  in comparison with the single  excitations when the driving field is on resonance with magnon mode, i.e., $\Delta/\kappa_m=0$.

In order to clarify the importance of the EP  and its effect on the quantum magnon and photon statistics, the second order correlation function $g_m^{(2)}(0)$ and $g_a^{(2)}(0)$, versus the normalized coupling strength $g/\kappa_m$, are respectively represented in Fig. (\ref{Fig69}) (a,b). As is mentioned above, $g/\kappa_m=1$ with $\kappa_a=-\kappa_m$ is not the only exceptional point of the system, $g=\kappa_a-\kappa_m$ with $\kappa_a \neq \kappa_m$ can also be considered as an exceptional point, which its influence are represented in Fig. (\ref{Fig69}).
In this figure, three different cases have been considered: (i) a coupled of decayed cavity and magnon mode with the same dissipation rates $\kappa_a=\kappa_m$, (ii) a coupled of dissipative magnon and amplified cavity modes with the same value of dissipation rates $\kappa_a=-\kappa_m$, and finally (iii) a coupled of dissipative magnon and amplified cavity modes with different values of dissipation rates $\kappa_a=-3\kappa_m$.
 To obtain these results, we have selected the resonance detuning, $\Delta/\kappa_m=0$. As is  evident, for $\kappa_a=\kappa_m$, $g_m^{(2)}(0)\to1$ and $g_a^{(2)}(0)\to1$  except that for small value of the coupling strength $0< g/\kappa_m<1$, which leads to a small antibunching. The behaviour of the two gain-loss coupled cavity magnon-modes is completely different. 
 By choosing $g/\kappa_m  $ and $\kappa_a=-\kappa_m$ or $g/\kappa_m \simeq 2$ and $\kappa_a=-3\kappa_m$ a strong magnon and photon antibunching and subsequently magnon and photon blockade can be achieved. As will be clarified in the next section, the strong antibunching in the $\mathcal{P}\mathcal{T}$-symmetric phase and EP is arising from the huge nonlinearity induced by the EP.

Finally, to see how the second order correlation functions, $g_m^{(2)}(0)$ and $g_a^{(2)}(0)$, explicitly depends on the magnon Kerr effect, which leads to a nonlinearity in the system as the main requirement for  blockade, $g_m^{(2)}(0)$ as well as $g_a^{(2)}(0)$ versus the nomalized nonlinearity $\chi/\kappa_m$ are represented in Fig. (\ref{Fig710})(a,b).
For $\chi=0$, $g_m^{(2)}(0)=g_a^{(2)}(0)=1$, which shows the Poissonian statistics  and means that  there is no magnon and photon blockade in the system. This result affirms that the nonlinear interaction is a necessary condition for the blockade effect. The results, which are obtained by selecting $g/\kappa_m\simeq2$ in two different configurations: (i) a coupled of dissipative cavity-magnon modes with different dissipation rates $\kappa_a=3\kappa_m$ and (ii) a coupled of dissipative magnon and amplified cavity mode with different dissipation rates $\kappa_a=-3\kappa_m$, show that an exceptional point can relax the requirement of strong Kerr nonlinearity such that even by a small magnon Kerr nonlinearity, strong magnon and photon excitation suppression is achieved. 
It worth emphasizing that in Figs. (\ref{Fig58}),  (\ref{Fig69}), (\ref{Fig710}), the numerical and analytical results have a hight agreement which shows the validity of the utilized approximation in Eq. (\ref{psi}). If we truncate the quantum state $\vert \psi\rangle$ over larger quantum number excitations, we can obtain more exact results with more consistency with the numerical ones.

\section{physical description of magnon blockade in the $\mathcal{P}\mathcal{T}$-symmetric phase}\label{sec4}

In order to obtain a physical intuition about the magnon and photon bunching suppression and hence magnon and photon blockade, we rewrite the Hamiltonian, Eq. (\ref{Hmatrix}), in terms of super mode operators. First, we define 
\begin{eqnarray}
    P\equiv \begin{pmatrix}
        \omega_a-i\kappa_a && g\\
        g && \omega_m-i\kappa_m
    \end{pmatrix},
\end{eqnarray}
which its eigenvalues are represented in Eq. (\ref{eigenvalues}). Moreover, we consider a $2\times2$ transformation matrix as $Q\equiv \begin{pmatrix}
    Q_{11}&&Q_{12}\\
    Q_{21} &&Q_{22}
\end{pmatrix}=\begin{pmatrix}
    \vert q_1\rangle &&  \vert q_2\rangle
\end{pmatrix}$ 
such that $PQ=QD$ with $D\equiv \begin{pmatrix}
    \omega_1 && 0\\
    0 && \omega_2
\end{pmatrix}$. 
Consequently, the Hamiltonian [\ref{Hmatrix}] can be rewritten as follows
\begin{eqnarray}
 \hat{H}&=&\begin{pmatrix}
     \hat{a}^{\dagger} && \hat{m}^{\dagger} 
 \end{pmatrix}   Q Q^{-1} P Q Q^{-1} 
 \begin{pmatrix}
     \hat{a}\\
     \hat{m}
 \end{pmatrix}\nonumber\\
 &=& \begin{pmatrix}
     \hat{A}^{\dagger} && \hat{M}^{\dagger}
 \end{pmatrix}
 \begin{pmatrix}
     \omega_1 && 0\\
     0 && \omega_2 
 \end{pmatrix}
 \begin{pmatrix}
     \hat{A}\\
     \hat{M}
 \end{pmatrix},
\end{eqnarray}
with the definition $\begin{pmatrix}
    \hat{A}\\
    \hat{M}
\end{pmatrix}= Q^{-1} \begin{pmatrix}
    \hat{a}\\
    \hat{m}
\end{pmatrix}$.
Now, we consider the nonlinear Kerr term $\hat{H}_{nl}=\hbar \chi (\hat{m}^{\dagger}\hat{m})^2$, in Eq. (\ref{rotated H}) and rewrite that in terms of the modified super mode operators $\hat{A}$ and $\hat{M}$. In more details, we substitute $\hat{m}=Q_{21}\hat{A}+Q_{22}\hat{M}$ in the nonlinear Kerr term and approximately obtain: $\hat{H}_{nl}\simeq \hbar \chi_1 \hat{A}^{\dagger}\hat{A}\hat{A}^{\dagger}\hat{A}+\hbar \chi_2 \hat{M}^{\dagger}\hat{M}\hat{M}^{\dagger}\hat{M} +\hbar \chi_3 \hat{A}^{\dagger}\hat{A}\hat{M}^{\dagger}\hat{M}$ in which the nonlinear coefficients are given by
\begin{eqnarray}
    \chi_1&=& \chi \vert Q_{21}\vert ^4,\\
    \chi_2&=& \chi \vert Q_{22}\vert ^4,\\
    \chi_3&=& 4\chi \vert Q_{21}\vert ^2 \vert Q_{22}\vert ^2.
\end{eqnarray}
By considering the $\mathcal{P}\mathcal{T}$-symmetric conditions in the proposed scheme, i.e., $\omega_a=\omega_m$ and $\kappa_m=-\kappa_a$, one can obtain the nonlinear coefficient as $\chi_j \propto \chi g^4 /(g^2-\kappa_a^2+\epsilon^2)^2$ with $j=1,2,3$ and $\epsilon$ is a small quantity originating from the Kerr nonlinearity. Obviously, in the vicinity of the EP, $g\simeq \kappa_m$, which also is known as the $\mathcal{P}\mathcal{T}$-phase, the effective Kerr nonlinearity of both the magnonic and  photonic modes is strongly enhanced. This interpretation denotes that even by selecting a small Kerr nonlinearity  $(\chi<<1)$, the magnon Kerr nonlinearity can be strikingly enhanced and a strong photonic Kerr nonlinearity can be generated.

\section{Experimental realization of the proposed scheme}\label{sec5}

The main characteristic of the proposed scheme is the $\mathcal{P}\mathcal{T}$-symmetry in which by properly opting the system parameters a perfect magnon and photon blockade can be achieved. Experimentally controlling a cavity magnonic system with $\mathcal{P}\mathcal{T}$-symmetric properties and the observation of the exceptional point has been reported in \cite{observation of pt-magnonic}. One of the requirements to achieve a $\mathcal{P}\mathcal{T}$-symmetric system is $\omega_a=\omega_m$, in which the magnon mode frequency is determined by $\omega_m=\gamma {B_0} +\omega_{m,ai}$, with $\gamma$ being the electron gyromagnetic ratio, $B_0$ representing the static magnetic field  and $\omega_{ai}$ denoting the anisotropic field. Evidently, this requirement can be easily realized by tunning $\omega_m$ via the static magnetic field which is utilized to excite the Kittel mode \cite{observation of pt-magnonic}. Another requirement is to satisfy $\kappa_a=-\kappa_m$, in which the cavity mode dissipation rate includes the extrinsic and intrinsic decay rates of the cavity, i.e., $\kappa_a=\kappa_a^{(int)}+\kappa_a^{(ext)}$ while the mognon mode dissipation rate $\kappa_m$ comes from the surface roughness as well as the impurities  and defects in the YIG sphere \cite{observation of pt-magnonic}. In the reported experimental scheme \cite{observation of pt-magnonic} the extrinsic  cavity mode dissipation rate is tailored to meet the mention requirement. Moreover, to observe the exceptional poitn, where $g=(\kappa_a-\kappa_m)$ the cavity-magnon coupling strength can be tunned by adjusting the displacement of the YIG sphere through which the amplitude of the magnetic field would be changed.
Finally, as the value of the Kerr coefficient strongly depends on the angle between the external magnetic field  and the crystalline axes, the small required Kerr nonlinearity can be achieved by choosing a suitable value of this angle \cite{magnon kerr observation}.

While there is no experimental techniques to directly distinguish the second order correlatin function of magnons, $g_m^{(2)}(0)$, as the quantum excitations of magnetically ordered systems in the solid YIG sphere, the detection of emitted photons from the cavity and subsequently measurement of the photonic second order correlation function, $g_a^{(2)}(0)$, would be performed by the Hanbury-Brown-Twiss inteferometer \cite{photon blockade CQED 1}. On the other hand, paying attention to the simultaneous behaviour of the second order magnon and photon correlation functions in the proposed scheme, the measurement of the photonic second order correlation function, $g_a^{(2)}(0)$, can be considered as a criteria not only for photon but also for magnon blockade.

 \section{summary, conclusion and outlooks}\label{sec6}

 In the present contribution, inspired by the effect of the $\mathcal{P}\mathcal{T}$-symmetry phase in the photonic devices, we proposed to study the behaviour of the magnon and photon statistics and hence magnon and photon blockade effect in such a case. To this end, we considered cavity-magnonic system in which by  properly controlling the cavity and magnon mode dissipation rates as well as the cavity-magnon coupling strength the $\mathcal{P}\mathcal{T}$-symmetric phase can be provided. We show that the $\mathcal{P}\mathcal{T}$-symmetry phase and subsequently the EP point can strongly suppress the magnon and photon bunching and consequently leads to perfect magnon and photon blockade even in the presence of very weak magnon Kerr nonlinearity in the system. In fact, the $\mathcal{P}\mathcal{T}$-symmetry phase can relax the large nonlinearity requirement which is considered as a necessary condition in conventional blockade. Although there is no photonic nonlinearity in the proposed scheme, a perfect photon blockade can occur simultaneously with magnon blockade. Thanks to this simultaneous behaviour of magnon and photon statistics, photon statistics can be experimentally considered as a  criteria of the magnon blockade in the proposed scheme.

 It worth emphasizing that although in this work we have considered magnon mode driving filed, the same results can be obtained by driving the cavity mode.

\end{document}